\documentclass[11pt,english,a4paper,fullpage]{article}
\usepackage{ae}
\usepackage[T1]{fontenc} 
\usepackage[utf8]{inputenc} 
\usepackage{graphicx,color} 
\usepackage{amssymb,amsmath,amsthm} 
\usepackage{float} 
\usepackage{multicol} 
\usepackage{appendix}
\DeclareGraphicsExtensions{.jpg}
\providecommand{\openone}{\leavevmode\hbox{\small1\kern-3.8pt\normalsize1}} 

\providecommand{\ket}[1]{|#1\rangle}

\providecommand{\ketbra}[2]{|#1\rangle\kern-2.8pt\langle#2|}
\providecommand{\sg}{\text{\,sg}}

\numberwithin{equation}{section}

\newtheorem{result}{Result}

\begin{document}
\title{Simulation of equatorial von Neumann measurements on GHZ states \\ using nonlocal resources}
\author{Jean-Daniel Bancal, Cyril Branciard, Nicolas Gisin \\
{\it Group of Applied Physics, University of Geneva,} \\
{\it 20 rue de l'Ecole-de-M\'edecine, CH-1211 Geneva 4, Switzerland}}
\date{November 2, 2010}
\maketitle

\abstract{Reproducing with elementary resources the correlations that arise when a quantum system is measured (quantum state simulation), allows one to get insight on the operational and computational power of quantum correlations. We propose a family of models that can simulate von Neumann measurements in the $x-y$ plane of the Bloch sphere on $n$-partite GHZ states using only bipartite nonlocal boxes. For the tripartite and fourpartite states, the models use only bipartite nonlocal boxes; they can be translated into classical communication schemes with finite average communication cost.}

\section{Introduction}

Understanding the nonlocal correlations created upon measurement of some entangled quantum system is a problem which runs up against our common representation of the world, by the very definition of nonlocality, i.e. violation of a Bell inequality \cite{BellB}. Indeed, no explanation one would reasonably accept as possible, like agreement prior to measurement, or subluminal communication of inputs, seems to be used by nature in order to create these correlations (see the numerous experimental violations of Bell inequalities \cite{Aspect99}).

Still, some insight on the power of such correlations was gained when people came out with models able to reproduce them in terms of classical resources. For instance, Toner and Bacon \cite{Toner} showed how to simulate von Neumann measurements on a singlet state with one bit of communication. Such a result puts an upper bound on the required amount of nonlocal resources needed for the reproduction of singlet correlations; it guarantees also that the corresponding correlations are not a stronger resource of nonlocality than 1 bit of classical communication.

A different kind of resources that was also considered are the so-called nonlocal boxes \cite{Barrett04}: these are simple nonlocal correlations which don't allow signaling. Successful simulation schemes using nonlocal boxes as unique nonlocal resources include the simulation of the singlet \cite{CGMP} and of partially entangled two-qubit states \cite{Brunner08}.

Concerning multipartite systems, communication models reproducing Pauli measurements on $n$-partite GHZ or on graph states have also been proposed \cite{Tessier05,Barrett07}. For arbitrary possible measurements on the tripartite GHZ state, previous studies suggested that its simulation with bounded communication might be impossible, taking as an example correlations corresponding to measurements of this state in the $x-y$ plane of the Bloch sphere \cite{BroadBentTapp}. In this paper, we construct a model which analytically reproduces these equatorial correlations, and whose only nonlocal resources are Popescu-Rohrlich (PR) boxes \cite{Popescu94} and Millionaire boxes \cite{Yao82}. Thus a finite number of bipartite nonlocal boxes are proven to be sufficient to reproduce these genuinely tripartite nonlocal correlations. Note also that even though our model doesn't give an upper bound on the worst-case communication cost, it does provide a communication model with finite expected communication cost, simulating for instance the tripartite GHZ state with an average total of 10 bits of communication between the parties (c.f. Appendix B).

The paper is organized as follows: first, we recall the correlations of the GHZ state that we want to simulate. We then present a model for the 3-partite case, and generalize it to more parties. We discuss the construction and then conclude.

\section{GHZ correlations}
Consider the $n$-partite GHZ state
\begin{equation}
\ket{GHZ_n}=\frac{1}{\sqrt2}\left(\ket{00\ldots0}+\ket{11\ldots1}\right).
\end{equation}
Our goal is to reproduce the correlations which are obtained when von Neumann measurements are performed on this state, by using other non-local resources such as non-local boxes (possibly supplemented with shared randomness).

For $n=2$, the protocol presented in \cite{CGMP} for the singlet state allows one to reproduce the correlations for any measurement settings, using one PR box. Here we recall the definition of a PR box:

\bigskip

\noindent \textbf{PR box}. \textit{A Popescu-Rohrlich (PR) box is a non-local box that admits two bits $x,y\in\{0,1\}$ as inputs and produces locally random bits $a,b\in\{0,1\}$, which satisfy the binary relation}
\begin{equation}\label{eq:PRbox}
a+b = xy.
\end{equation}

\bigskip

Going to $n\geq3$, we shall only consider measurements in the $x-y$ plane (equatorial measurements), which have the nice feature of producing unbiased marginals: all correlation terms involving strictly fewer than $n$ parties vanish. We write each party's measurement operator as: $A=\cos{\phi_a}\,\sigma_X+\sin{\phi_a}\,\sigma_Y$, $B=\cos{\phi_b}\,\sigma_X+\sin{\phi_b}\,\sigma_Y, \ldots $ Denoting the binary result of each measurement by $\alpha, \beta, \ldots \in\{-1,1\}$, the correlations we are interested in are given by
\begin{equation}\label{eq:marg}
\langle\alpha\rangle=\langle\beta\rangle=\ldots=\langle\alpha\beta\rangle=\ldots=0
\end{equation}
for all sets of fewer than $n$ parties, and
\begin{equation}\label{eq:corr}
\langle \alpha\beta \ldots \omega \rangle = \cos(\phi_a+\phi_b+\ldots+\phi_z)
\end{equation}
for the full $n$-partite correlation term.
In other words, outcomes appear to be random except when all of them are considered together, in which case their correlation takes a form reminiscent of the singlet state. To simulate such correlations, nonlocal boxes similar to the Millionnaire box will be useful, so let us recall what a Millionnaire box is:

\bigskip

\noindent \textbf{M box}. \textit{A Millionaire box is a non-local box that admits two continuous inputs $x,y\in[0,1[$ and produces locally random bits $a,b\in\{0,1\}$, such that}
\begin{equation}\label{eq:Mbox}
a+b = \sg(x-y)
\end{equation}
where the sign function is defined as $\sg(x)= 0$ if $x > 0$ and $\sg(x)= 1$ if $x \leq 0$.

\bigskip

It is worth mentioning that even though we restrict the set of possible measurements on the GHZ states, the correlations we consider can still exhibit full $n$-partite nonlocality. Indeed, the Svetlichny inequality for $n$ parties can be maximally violated with settings in the $x-y$ plane \cite{Collins02,Seevinck02}. This implies that in order to simulate these correlations, any model must truly involve all $n$ parties together \cite{Bancal09}.

\section{Simulation model for the 3-partite GHZ state}
\begin{figure}
\begin{center}
\includegraphics[width=6cm]{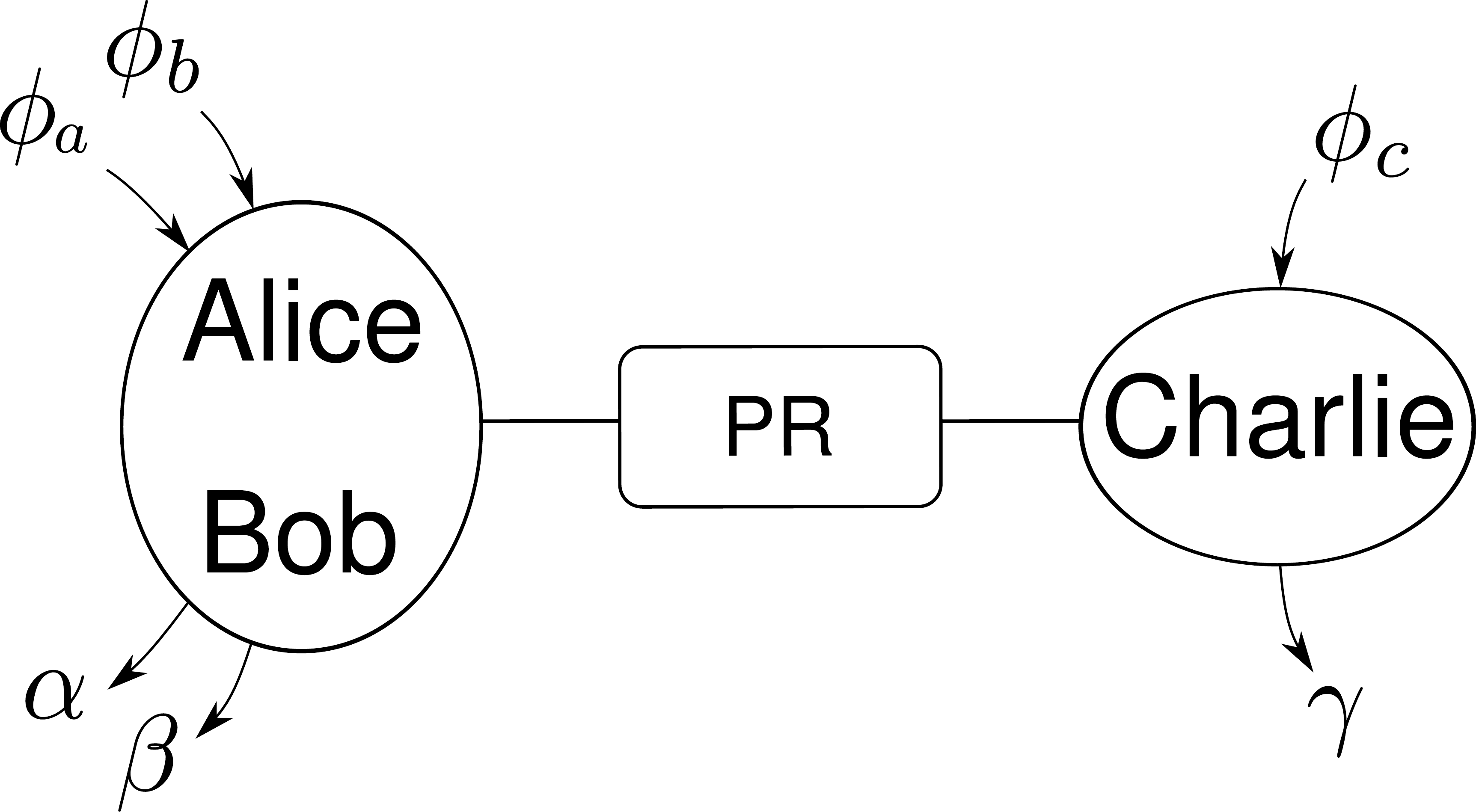}
\caption{Simulation of $\ket{GHZ_3}$ in a Svetlichny scenario: Alice and Bob form a group and can share their information with each other, while Charlie is separated from them. In this scenario 1 PR box allows one to reproduce the equatorial correlations.}
\label{fig:1PR}
\end{center}
\end{figure}
Let us consider the above correlations for $n=3$ parties, for which the outcomes of all parties need to be correlated according to $\langle\alpha\beta\gamma\rangle=\cos(\phi_a+\phi_b+\phi_c)$.

As a first step towards the simulation of these correlations, let us relax some of the constraints and allow two parties to cooperate in a Svetlichny-like scenario \cite{svetlichny} (see Figure~\ref{fig:1PR}): for instance Alice and Bob would be allowed to communicate with each other, but not with Charlie who is kept isolated from them. In such a scenario, the three parties could create correlations of the desired form with one PR box by using the protocol of \cite{CGMP} to generate outputs $\tilde{\alpha}$ and $\tilde{\gamma}$ that have a cosine correlation of the form $\langle\tilde{\alpha}\tilde{\gamma}\rangle=\cos(\phi_{ab}+\phi_c)$, with a fictitious measurement angle $\phi_{ab}=\phi_a+\phi_b$. By then setting either $\alpha=\tilde{\alpha},\beta=+1$ or $\alpha=-\tilde{\alpha},\beta=-1$ (each with probability 1/2), and $\gamma=\tilde{\gamma}$, they would recover the desired tripartite correlations $\langle\alpha\beta\gamma\rangle=\cos(\phi_a+\phi_b+\phi_c)$.

Of course, letting Alice and Bob share their inputs is not satisfactory yet, as this would require signaling between them. We shall now see that it is actually possible to re-separate them, while keeping the tripartite correlation term unchanged.

In order to do so, let us recall that the model used above to create the bipartite cosine correlation with a PR box works by asking the parties (here, Alice-Bob together and Charlie) to input in the box terms of the form \cite{CGMP}
\begin{equation}\label{eq:inputsPR}
\begin{split}
x&=\sg(\cos(\phi_{ab}-\varphi_1))+\sg(\cos(\phi_{ab}-\varphi_2))\\
z&=\sg(\cos(\phi_c-\varphi_+))+\sg(\cos(\phi_c-\varphi_-)),
\end{split}
\end{equation}
where $\varphi_1, \varphi_2, \varphi_+, \varphi_-$ are hidden variables shared by all parties (that we shall define later). One can see that in the Svetlichny scenario, Alice and Bob don't really need to share their measurement angles, but only the terms $\sg(\cos(\phi_a+\phi_b-\varphi_\lambda))$. Hopefully, there is a way for Alice and Bob to compute this function nonlocally by using the forementioned  Millionaire box (M box). For convenience, let us define the following nonlocal box:

\bigskip

\noindent \textbf{Cosine box}. \textit{A bipartite Cosine box (C box) is a non-local box that admits two angles $\phi_a,\phi_b\in[0,2\pi[$ as inputs and produces locally random binary outcomes $a,b\in\{0,1\}$, correlated according to}
\begin{equation}\label{eq:Cbox}
a+b = \sg(\cos(\phi_a+\phi_b)).
\end{equation}

\bigskip

We show in Appendix A that a bipartite C box is equivalent to a M box. C boxes are exactly what we need for our problem, as the following result shows:

\begin{result}
Equatorial von Neumann measurements on the tripartite GHZ state can be simulated with 2 C boxes and 2 PR boxes.
\end{result}

\begin{proof}
The simulation can be realized with the following model; we refer to Figure~\ref{fig:2C2PR} for the distribution of the non-local boxes between the three parties Alice, Bob and Charlie, and for the numbering of their inputs (denoted $x_i, y_i$ and $z_i$ for each party respectively) and outputs (denoted $a_i, b_i$ and $c_i$).

\begin{figure}
\begin{center}
\includegraphics[width=4.5cm]{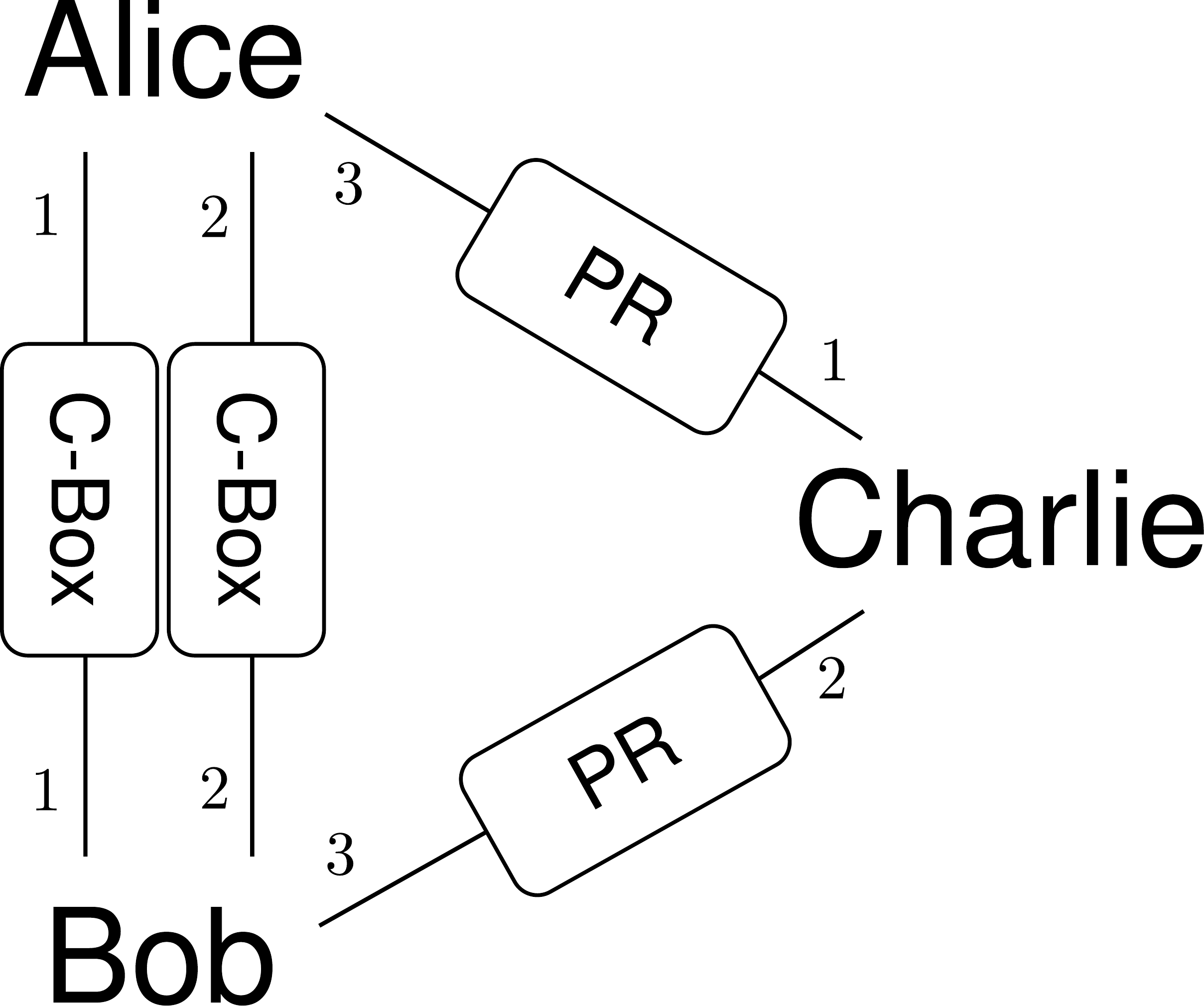}
\caption{Setup for the simulation of $\ket{GHZ_3}$ in the $x-y$ plane. The Alice-Bob group was split by using two C boxes and a second PR box.}
\label{fig:2C2PR}
\end{center}
\end{figure}

Let Bob and Charlie share two independent random vectors $\vec\lambda_1, \vec\lambda_2$ uniformly distributed on the sphere $S_2$. We define $\vec\lambda_\pm=\vec\lambda_1\pm\vec\lambda_2$ and refer to $\varphi_1, \varphi_2, \phi_+, \varphi_-$ for their phase angle in polar coordinates. Let the parties input the following variables into their boxes:
\begin{equation}
\begin{split}
x_1&=\phi_a,\ \ \ \ \ \ \ \ \ \, x_2=\phi_a,\ \ \ \ \ \ \ \ \ \, x_3=a_1+a_2\\
y_1&=\phi_b+\varphi_1, \ \ \ y_2=\phi_b+\varphi_2,\ \ \ y_3=b_1+b_2\\
z_1&=z_2=\sg(\cos(\phi_c-\varphi_+))+\sg(\cos(\phi_c-\varphi_-)).
\end{split}
\end{equation}
The three parties then output:
\begin{equation}
\begin{split}
\alpha &= (-1)^A,\ \text{with } A={a_1+a_3}\\
\beta &= (-1)^B,\ \text{with } B={b_1+b_3}\\
\gamma &= (-1)^C,\ \text{with } C={c_1+c_2+\sg(\cos(\phi_c-\varphi_+))}.
\end{split}
\end{equation}

The output of each party is the \textsc{xor} of outputs received from nonlocal boxes shared with all other parties. Since these boxes are no-signaling, a single output of any nonlocal box is necessarily random. The only way as not to get a random average correlation is thus to consider all parties together, since missing one produces a random term. All correlations involving fewer than 3 parties thus average to zero.

Concerning the 3-party correlations, we have:
\begin{equation}
\begin{split}
A+B+C&=(a_1+b_1)+(a_3+c_1)+(b_3+c_2)+\sg(\cos(\phi_c-\varphi_+))\\
&=(a_1+b_1)+x_3z_1+y_3z_2+\sg(\cos(\phi_c-\varphi_+))\\
&=(a_1+b_1)+(a_1+b_1+a_2+b_2)z_1+\sg(\cos(\phi_c-\varphi_+))\\
&=(a_1+b_1)+\sg(\cos(\phi_c-\varphi_+))\\
& \qquad +\left[\sg(\cos(\phi_a+\phi_b+\varphi_1))+\sg(\cos(\phi_a+\phi_b+\varphi_2))\right] \\
& \qquad \qquad \qquad \qquad \times \left[\sg(\cos(\phi_c-\varphi_+))+\sg(\cos(\phi_c-\varphi_-))\right]\\
&=\sg(\vec v_{ab} \cdot \vec \lambda_1)+\sg(\vec c \cdot \vec \lambda_+)\\
& \qquad +\left[\sg(\vec v_{ab} \cdot \vec \lambda_1)+\sg(\vec v_{ab} \cdot \vec \lambda_2)\right]\left[\sg(\vec c \cdot \vec \lambda_+)+\sg(\vec c \cdot \vec \lambda_-)\right]
\end{split} \label{eq:ABC}
\end{equation}
where we defined $\vec v_{ab}=(\cos(-\phi_a-\phi_b),\sin(-\phi_a-\phi_b),0)$ and with $\vec c = (\cos\phi_c,\sin\phi_c,0)$ being Charlie's setting. Following the proof of \cite{CGMP} (see also \cite{Toner}), we find that the average of this quantity over the values of the hidden variables $\vec \lambda_1$ and $\vec \lambda_2$ is:
\begin{equation}
\langle A+B+C \rangle = \frac{1-\vec v_{ab}\cdot\vec c}2 = \frac{1-\cos(\phi_a+\phi_b+\phi_c)}2
\end{equation}
which leads, as requested, to
\begin{equation}
\langle\alpha\beta\gamma\rangle=\cos(\phi_a+\phi_b+\phi_c).
\end{equation}
\end{proof}

Coming back to the Svetlichny construction we see that it was indeed possible to split the Alice-Bob group by allowing them to share two C boxes. Concerning the PR box, it had to be split also, into two new PR boxes, in order to recover the desired result: the computation made by the PR box in the Svetlichny setup is now performed nonlocally, by the 2 PR boxes, using inputs distributed over the 3 parties.

We restricted here to measurements in the $x-y$ plane, but with a slight modification, Charlie could actually simulate any measurement basis. Indeed a way to understand the appearance of the model for the singlet state (the bipartite cosine correlation), in the Svetlichny scenario, is to realize that the fictitious measurement angle $\phi_{ab}=\phi_a+\phi_b$ that Alice and Bob used above corresponds to the direction in which they would prepare a state for Charlie if they were to measure their part of the original GHZ state in their respective bases. In other words, in the quantum scenario, when Alice and Bob measure the GHZ state, they prepare one of the two state
\begin{equation}
\ket{z_{\pm}}=\frac{1}{\sqrt2}(\ket{0}\pm e^{-i(\phi_a+\phi_b)}\ket{1})
\end{equation}
for Charlie. But a way for them to prepare one of these states if they share a singlet (or rather, a bipartite GHZ or $\ket{\Phi^+}$ state) with Charlie, is by measuring their part of the $\ket{\Phi^+}$ state along $\phi_{ab}$, which is what they effectively do in our model. So in fact they prepare a state $\ket{z_\pm}$ for Charlie, which he can measure in the direction he wants (in particular, outside the $x-y$ plane). The only modification in the model needed for that is that Charlie should use $z_1=z_2=\sg(\vec c \cdot \vec \lambda_+)+\sg(\vec c \cdot \vec \lambda_-)$ and $C=c_1+c_2+\sg(\vec c \cdot \vec \lambda_+)$ to allow his measurement to point outside the $x-y$ plane.

\bigskip

We do not claim that the above model is optimal. It could be that strictly fewer nonlocal boxes are actually enough to reproduce the same correlations. It is nonetheless remarkable that truly tripartite correlations can be simulated with bipartite nonlocal resources only.

It is also quite surprising that the model we presented here does not need more shared randomness than in the bipartite case. It might possibly be that a model that would use fewer nonlocal resources would require more shared randomness.

\section{Simulation model for the 4-partite GHZ state}
In the previous section we showed how to split $\phi_{ab}$ from equation \eqref{eq:inputsPR} into two phases $\phi_a,\phi_b$, in order to re-separate the group formed by Alice and Bob in the Svetlichny scenario. It is in fact similarly possible to split $\phi_c$ in order to have a total of 4 parties into play:

\begin{result}
Equatorial von Neumann measurements on the 4-partite GHZ state can be simulated with 4 C boxes and 4 PR boxes.
\end{result}

\begin{proof}
The simulation can be realized with the following model, analogous to the previous one; we now refer to Figure~\ref{fig:4C4PR} for the distribution of the non-local boxes between the four parties Alice, Bob, Charlie and Dave, and for the numbering of their inputs ($x_i, y_i, z_i$ and $w_i$) and outputs ($a_i, b_i, c_i$ and $d_i$).

\begin{figure}
\begin{center}
\includegraphics[width=5cm]{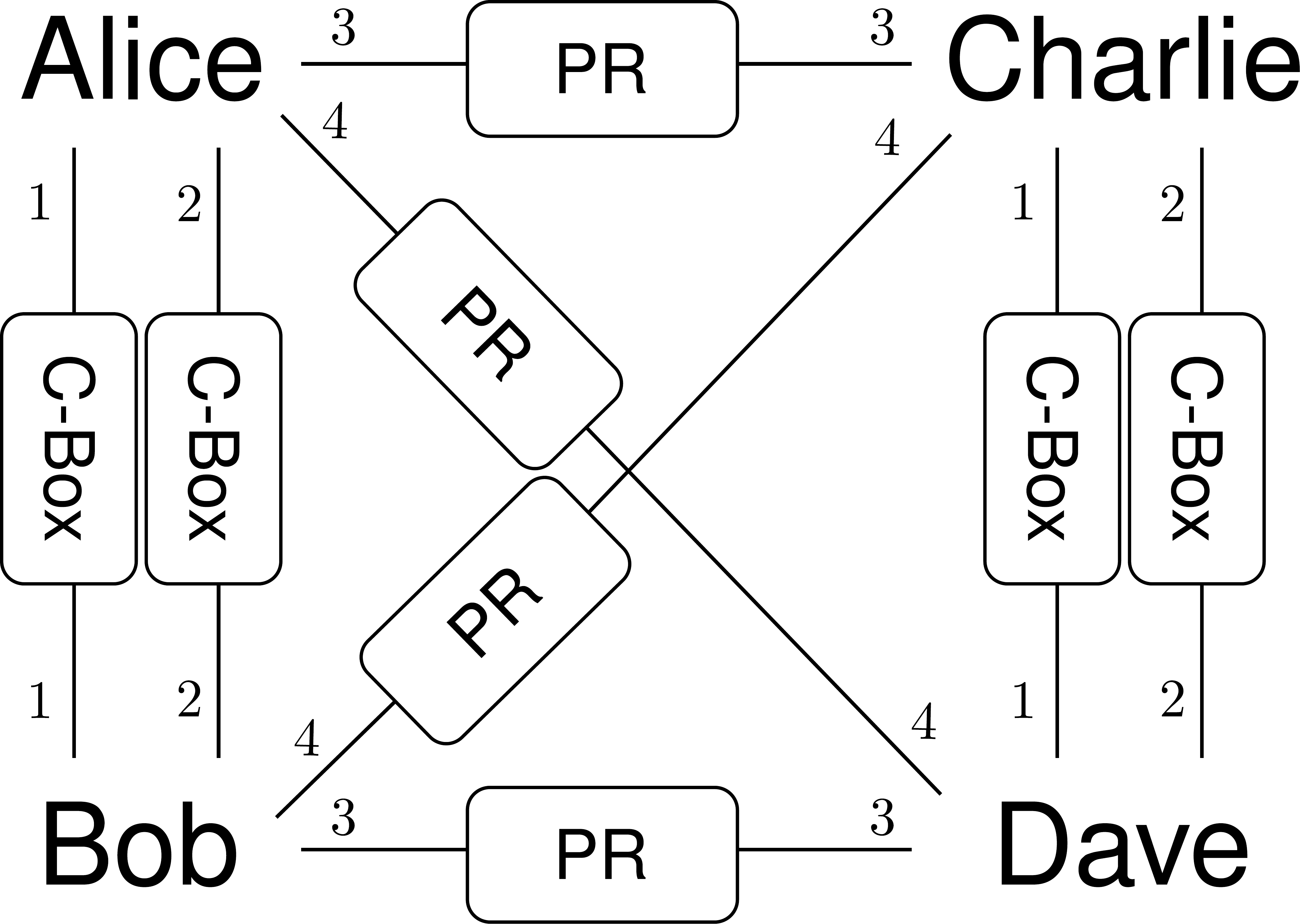}
\caption{Setup for the simulation of $\ket{GHZ_4}$ in the $x-y$ plane. Notice that if any of the 4 parties is taken away (together with the non-local boxes it shares with the other parties), we recover a setup with 2 PR and 2 C boxes, which corresponds to the simulation setup for $\ket{GHZ_3}$, as in Figure~\ref{fig:2C2PR}.}
\label{fig:4C4PR}
\end{center}
\end{figure}

Bob and Charlie still share two independent random vectors $\vec\lambda_1, \vec\lambda_2$ uniformly distributed on the sphere $S_2$. With the same notations as before, let the four parties now input the following variables into their boxes:
\begin{align}
x_1&=\phi_a, &x_2&=\phi_a, &x_3&=x_4=a_1+a_2\nonumber\\
y_1&=\phi_b+\varphi_1, &y_2&=\phi_b+\varphi_2, &y_3&=y_4=b_1+b_2\nonumber\\
z_1&=\phi_c-\varphi_+, &z_2&=\phi_c-\varphi_-, &z_3&=z_4=c_1+c_2\\
w_1&=\phi_d, &w_2&=\phi_d, &w_3&=w_4=d_1+d_2\nonumber
\end{align}
The parties should then output:
\begin{equation}
\begin{split}
\alpha &= (-1)^A,\ \text{with } A={a_1+a_3+a_4}\\
\beta &= (-1)^B,\ \text{with } B={b_1+b_3+b_4}\\
\gamma &= (-1)^C,\ \text{with } C={c_1+c_3+c_4}\\
\delta &= (-1)^D,\ \text{with } D={d_1+d_3+d_4}
\end{split}
\end{equation}

For the same reason as in the tripartite case, all correlations of fewer than four parties vanish. For the 4-partite correlation term, the calculation of $A+B+C+D$ is straightforward, following similar lines as in the tripartite case. It leads to a similar expression as in (\ref{eq:ABC}), except that $\vec c$ should now be replaced by $\vec v_{cd}=(\cos(\phi_c+\phi_d),\sin(\phi_c+\phi_d),0)$. This leads to the requested 4-partite correlation term:
\begin{equation}
\langle\alpha\beta\gamma\delta\rangle=\cos(\phi_a+\phi_b+\phi_c+\phi_d).
\end{equation}


\end{proof}


\bigskip

Again, there is no claim of optimality for the above model, but it is also remarkable that truly 4-partite correlations can still be simulated with bipartite nonlocal resources only, and no more shared randomness than for the bipartite case.

\section{Going to more parties}

\subsection{Possible extension of the model to any number of parties}
In the last two sections, we showed how to construct models for the simulation of GHZ states involving $n=3,4$ parties by splitting the $n$ parties into two groups. Each group then had to calculate functions of the form $\sg(\cos(\Sigma\phi_{a_i}+\varphi_\lambda))$ with for instance $\Sigma\phi_{a_i}=\phi_a+\phi_b$, $\varphi_\lambda=\varphi_1$. Now, if we consider more parties, splitting them into two groups necessarily results in at least one of the groups having more than two parties. One could for instance have $n-1$ parties on one side and $1$ party on the other side. The sign function that each group has to calculate thus involves in general more than two phase angles. This motivates the definition of a generalization of the C box to $n$ parties:

\bigskip

\noindent \textbf{Multipartite Cosine box}. \textit{An $n$-partite C box is a non-local non-signaling box that admits $n$ angles $\phi_i\in[0,2\pi[$ as inputs and produces binary outcomes $a_i\in\{0,1\}$, correlated according to}
\begin{equation}
\sum_i a_i = \sg(\cos(\sum_i \phi_i)).
\end{equation}
\textit{The outcomes of the box are locally random. Also, all correlations involving fewer than $n$ outputs vanish.}

\bigskip

Multipartite C boxes allow one to generalize our model to the simulation of multipartite GHZ state with any number of parties, by separating the $n$ parties into two groups, consisting of $k$ parties on one side and $n-k$ parties on the other:

\begin{result}
Equatorial von Neumann measurements on $n$-partite GHZ states can be simulated with 2 $k$-partite C boxes + 2 $(n-k)$-partite C boxes + $k(n-k)$ PR boxes (for any $0<k<n$).
\end{result}

\begin{proof}[Sketch of the proof]

Following the previous constructions, the group with $k$ parties needs to calculate nonlocally two terms of the form $\sg(\cos(\phi_1+\ldots+\phi_k+\varphi_\lambda))$, which can be done by using two $k$-partite C boxes, and the other group can similarly do its job with two $(n-k)$-partite C boxes. As it was the case for the 4-partite case, each party from the first group also needs to share a PR box with each other party in the second group. We thus understand that by separating the $n$ parties into these two groups, a total of 2 $k$-partite C boxes + 2 $(n-k)$-partite C boxes + $k(n-k)$ PR boxes is sufficient to simulate the correlations of the $n$-partite GHZ state measured in the $x-y$ plane. Interestingly again, no more shared randomness than for the bipartite case is required.

\end{proof}

\subsection{A simpler model}
If we allow the parties to share nonlocal boxes involving more than two parties, then there is actually a simpler model which uses a single $n$-partite C box to reproduce the equatorial GHZ correlations (as defined by \eqref{eq:marg} and \eqref{eq:corr}):

\begin{result}
Equatorial von Neumann measurements on $n$-partite GHZ states can be simulated with a single $n$-partite C box.
\end{result}

\begin{proof}
Consider indeed the following strategy: Alice generates a random variable $\varphi_\lambda \in [-\pi/2,\pi/2]$ according to the distribution $\rho(\varphi_\lambda) = \frac12\cos(\varphi_\lambda)$. She inputs $\phi_a+\varphi_\lambda$ in the $n$-partite C box, while all other $n-1$ partners simply input their measurement angle. From the outputs $a,b,\ldots$ of the box, each party can compute the final outputs $\alpha = (-1)^a, \beta = (-1)^b, \ldots$ All correlations between the outputs that involve fewer than $n$ parties vanish, while for the $n$-partite correlation term, they get, as requested:
\begin{equation}
\begin{split}
\langle\alpha\beta\ldots\omega\rangle&=\int_{-\pi/2}^{\pi/2}(-1)^{\sg(\cos(\varphi_\lambda+\phi_a+\phi_b+\ldots+\phi_z))}\rho(\varphi_\lambda)d\varphi_\lambda\\
&=
\begin{cases}
\frac12 \int_{-\pi/2}^{\pi/2-\Sigma\phi_i} \cos\varphi_\lambda \, d\varphi_\lambda - \frac12 \int_{\pi/2-\Sigma\phi_i}^{\pi/2} \cos\varphi_\lambda \, d\varphi_\lambda & \text{if } 0<\Sigma\phi_i<\pi\\
-\frac12 \int_{-\pi/2}^{-\pi/2-\Sigma\phi_i} \cos\varphi_\lambda \, d\varphi_\lambda + \frac12 \int_{-\pi/2-\Sigma\phi_i}^{\pi/2} \cos\varphi_\lambda \, d\varphi_\lambda & \text{else}
\end{cases}\\
&=\cos(\phi_a+\phi_b+\ldots+\phi_z)
\end{split}
\end{equation}

\end{proof}

Note that in the bipartite case, this model gives a new, simplified, way of simulating the equatorial correlations of the singlet state with a single Millionaire box. It is worth noting that it does not require any shared randomness. It uses however a strictly stronger non-local resource than the model with one PR box \cite{CGMP}, since an M box cannot be simulated with one PR box (c.f. Appendix B). 

\bigskip

Compared to this last simple model, our previous construction allows one to reduce the multipartiteness of the nonlocal boxes used to simulate the same correlations. Finitely many nonlocal boxes involving no more than $\lceil n/2 \rceil$ parties are sufficient to reproduce $n$-partite equatorial GHZ correlations. In particular, for $n\leq 4$, bipartite resources are sufficient.

If one really wants to use only bipartite nonlocal boxes, we show in Appendix C that multipartite nonlocal boxes with continuous inputs, binary outputs, and only fully $n$-partite non-vanishing correlations, can always be simulated with bipartite boxes, as it is the case for boxes with a finite number of inputs \cite{Barrett05}. However, the construction we use is quite special, as the boxes we need can have inputs or outputs that cannot be written as real numbers.

\section{Conclusion}

We proposed models reproducing the correlations of the tripartite and 4-partite GHZ states measured in the $x-y$ plane, with a finite number of bipartite nonlocal boxes. Extending our results to $n$-partite GHZ states was possible after releasing the requirement that the nonlocal boxes had to be bipartite.


We believe that our results give a new motivation for finding whether or not the GHZ correlations can also be simulated in a bounded communication scheme. Note that our models can be translated into finite expected communication schemes, since a PR box can be replaced by 1 bit of communication and an M box (bipartite C box) by 4 bits in average, as we show in Appendix B. This gives a model with an average of 10 bits of communication between the parties. Note that this model with finite expected communication could also be recast as a detection loophole model.

More generally, it would be interesting to know whether the simulation of $n$-partite GHZ states can always be achieved with a finite amount of bipartite resources only (for instance a finite number of M boxes). Considering also measurements outside of the $x-y$ plane seems quite challenging because the marginals don't vanish anymore, but it would certainly be of interest too.

Finally, it would be worth studying other multipartite quantum correlations. The W state for instance, seems to be a good candidate for this, when measurements are again restricted to the $x-y$ plane, because of the simplicity of its correlations. Indeed they only consist of bipartite correlation terms of the form $\langle\alpha\beta\rangle=\frac{2}{n} \cos(\phi_a-\phi_b)$, all other correlation terms being 0 for any number of parties $n$ \footnote{It is not known whether these correlations are nonlocal for all $n$, but in the case $n=4$ there exists a Bell inequality which allows one to show that these correlations are indeed nonlocal \cite{Bancal09b}.}.

\section{Acknowledgments}

We thank Stefano Pironio for useful discussions.
We acknowledge support by the Swiss NCCR Quantum Photonics and the European ERC-AG QORE.

\bigskip

\subsection*{Appendix A. The bipartite Cosine box is equivalent to a Millionaire box}

Here we show that in the bipartite case, the Cosine box is equivalent to a Millionaire box, up to local operations on the inputs and outputs. The general $n$-partite C box can thus also somehow be seen as a generalization of a M box to more parties.

\bigskip

Let us first give the intuition. It is indeed clear that a bipartite C box is equivalent to a ``sine box", that would take two angles $\phi_a,\phi_b\in[-\pi,\pi[$ as inputs and would output two locally random bits $a,b\in\{0,1\}$ with correlations satisfying
\begin{equation}\label{eq:Sbox}
a+b=\sg(\sin(\phi_a-\phi_b)).
\end{equation}
Now, if $\phi_a\in[-\pi,0[$, Alice can input $\phi_a+\pi\in[0,\pi[$ in the sine box instead of $\phi_a$, and flip her output so that (\ref{eq:Sbox}) is still satisfied. This also holds for Bob; we can thus assume that $\phi_a,\phi_b\in[0,\pi[$. In that case, $\sg(\sin(\phi_a-\phi_b)) = \sg(\phi_a-\phi_b)$. The sine box thus compares the values of the two real numbers $\phi_a,\phi_b$; this is exactly what a M box would do!

\bigskip

\begin{figure}
\begin{center}
\includegraphics[width=9cm]{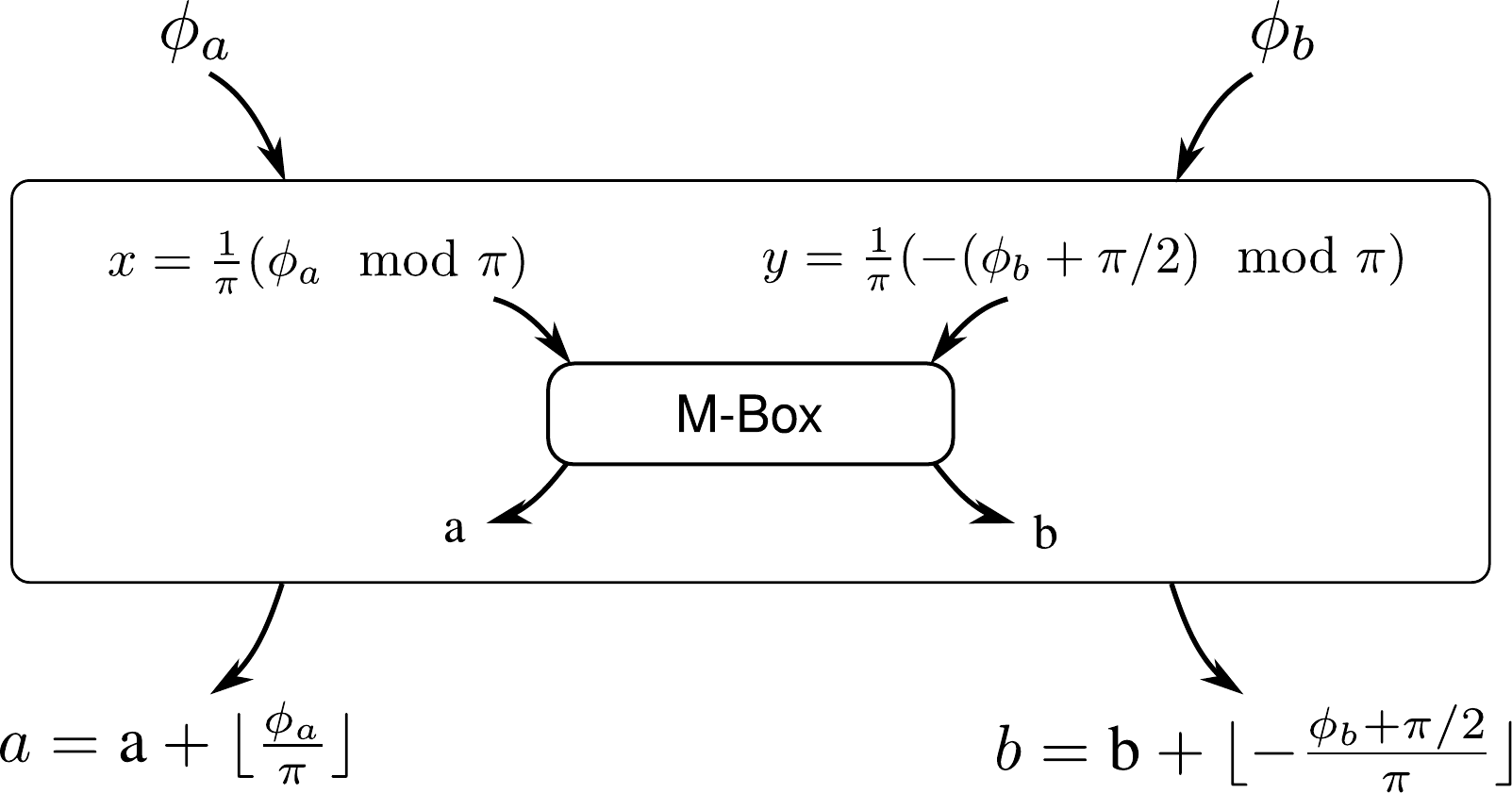}
\caption{How to realize a bipartite Cosine box from a Millionaire box.}
\label{fig:Cbox}
\end{center}
\end{figure}

More precisely, to construct a C box from a M box \eqref{eq:Mbox}, Alice and Bob can input $x=\frac1\pi(\phi_a \mod \pi)$ and $y=\frac1\pi(-(\phi_b+\pi/2) \mod \pi)$. From the outputs $\text{a}$ and $\text{b}$ of the M box, they can calculate $a=\text{a}+\lfloor\frac{\phi_a}\pi\rfloor$ and $b=\text{b}+\lfloor-\frac{\phi_b+\pi/2}\pi\rfloor$, which satisfy $a+b=\sg(\cos(\phi_a+\phi_b))$, as requested. This construction is illustrated on Figure~\ref{fig:Cbox}.

Reciprocally, the M box can trivially be reproduced with a C box, if Alice inputs $\phi_a=x$ and Bob inputs $\phi_b=-y-\pi/2$.



\subsection*{Appendix B. Expected communication cost of a Millionaire box}

We show in this Appendix that a Millionaire box cannot be simulated with finite communication. We propose however a scheme to simulate it with 2-way communication, unbounded in the worst case, but with a finite expected number of bits.

\bigskip

Suppose first that Alice and Bob have a finite number $2^k$ of possible inputs ($k$ bits). We show with a crossing sequence argument~\cite{kushilevitz_nisan} that any communication scheme that can simulate the outputs of a M box for this number of possible inputs necessarily uses at least $k$ bits. This shows that in the limit of infinitely many inputs (ie, for the general M box), unbounded communication is required.

Indeed suppose that a scheme using $k_0<k$ bits of (possibly 2-way) communication can simulate the M box with $2^k$ possible inputs on each side. In particular it can simulate it when Alice and Bob use the same inputs $x=y$. But since the number of all possible identical inputs ($2^k$) is greater than the number of possible messages exchanged by Alice and Bob during the communication procedure ($2^{k_0}$), there must be at least two different pairs of identical inputs $x_0=y_0$ and $x_1=y_1$ (with $x_0\neq x_1$) for which the communication pattern is the same. This communication pattern is then also the same if the inputs are $x=x_0$, $y=y_1$ or if $x=x_1$, $y=y_0$, because Alice and Bob will not see any difference. So if the simulation of the M box produces outputs saying that $x_0\leq y_1=x_1$, it will also say that $x_1\leq y_0=x_0$, which contradicts the fact that $x_0\neq x_1$.

Thus simulation of a M box with $2^k$ possible inputs on each side necessarily needs at least $k$ bits to be exchanged between the parties. So in the limit $k\to\infty$, the required amount of exchanged bits goes to infinity.




\bigskip

Here is however a simple model that uses a finite average of 4 bits of communication (2 bits in each direction) to simulate a M box.

Let us write the two inputs $x,y \in [0,1]$ of the M box in basis 2, so that each digit is either a 0 or a 1. Consider the following protocol, starting with $n=1$:
\begin{enumerate}
 \item Alice sends her $n^{\text{th}}$ digit to Bob.
 \item Bob compares the bit he received with his $n^{\text{th}}$ digit and answers 0 if they are the same and 1 if they are different.
 \item If Alice receives a 0, she iterates $n$ and goes back to step number 1. If however she receives a 1, then they both know which one of them has the largest input number. Alice can output a predetermined random bit, and Bob a bit correctly correlated to Alice's, so as to reproduce the behavior of the M box.
\end{enumerate}

The average number of communication cycles needed in this scheme depends on the probabilistic distribution of $x$ and $y$. In particular if these distributions are independent and uniform on the interval $[0,1]$, the probability that the protocol stops at the $n^{\text{th}}$ step is $2^{-n}$, and therefore the expected number of rounds is
\begin{equation}
\sum_{n=1}^{\infty}n\ 2^{-n}=\frac{1/2}{(1-1/2)^2}=2.
\end{equation}
Since each round uses 2 bits of communication (one in each direction), a total of 4 bits of communication is needed on average.

Similar ideas can also be used to simulate $n$-partite C boxes with finite expected communication.

\subsection*{Appendix C. Simulation of $n$-partite correlations with bipartite nonlocal resources}

Consider an $n$-partite probability distribution for continuous inputs $x_i\in\mathbb{R}$ and binary outputs $a_i=\{0,1\}$, which contains vanishing correlations for up to $n-1$ parties. We show that it can be simulated with only bipartite nonlocal boxes. This can be seen as a generalization of \cite{Barrett05}, in which a similar decomposition was constructed for distributions with finitely many inputs in terms of PR boxes.

\bigskip

To show this result it is sufficient to concentrate on the total correlation term $\sum_i a_i = f(x_i)$ involving all parties' outputs $a_i$, because all other correlation terms can then be put to zero by letting all pairs of parties decide randomly to permute their output or not.

Consider thus the $n$-partite correlation term. We proceed by recursion: starting with the case $n=2$, in which it is obvious that any bipartite no-signaling correlation can be produced by a bipartite nonlocal box satisfying
\begin{equation}\label{eq:box1}
a_1+a_2=f(x_1,x_2).
\end{equation}

Now let us suppose that we have a model which can reproduce any correlation term for $n-1$ parties. Any $n$-partite correlation term can then be simulated the following way: for each value $z$ that the $n^{\text{th}}$ party's input $x_n$ can take, define the following function for the $n-1$ first parties: $f_z(x_1,\ldots,x_{n-1})\equiv f(x_1,\ldots,x_{n-1},z)$. Each of these functions can be implemented by the scheme reproducing the ($n-1$)-partite correlation functions. Now each of the $n-1$ parties can collect the outputs $\alpha_i(z)$ it received for each possible value of $z$, and plug them into a special kind of bipartite nonlocal box it shares with the last party. This box takes as inputs a function $\mathfrak{f}:[0,1]\to\{0,1\}$ on one side (i.e. a continuous number of binary inputs) and a real number $x\in[0,1]$ on the other side, and produces binary outputs $a$ and $b$, such that
\begin{equation}\label{eq:box2}
a+b=\mathfrak{f}(x)
\end{equation}

If all parties input $\mathfrak{f}=\alpha_i(z)$ into such a box they share with the last party, and this last party inputs $x_n$ into all of these boxes, writing the outputs of each of these boxes $a_i$ and $a_n^i$, we can set the last party's output to be $a_n=\sum_{i=1}^{n-1}a_n^i$ to get a correlation term:
\begin{equation}
\begin{split}
\sum_{i=1}^n a_i &= \sum_{i=1}^{n-1} a_i + a_n = \sum_{i=1}^{n-1} (a_i + a_n^i)= \sum_{i=1}^{n-1} \alpha_i(x_n)\\
&= f_{x_n}(x_1,\ldots,x_{n-1})= f(x_1,\ldots,x_n)
\end{split}
\end{equation}
as required.

Note that this construction needs continuously many nonlocal boxes. To avoid that, one could collect all the boxes \eqref{eq:box1} that calculate $f_z(x_i,x_j)$ for all $z$ into a single one that would output all the values at the same time. Note however that such a box would actually output continuous outputs of cardinality $\aleph_2$ (i.e. binary functions defined on $\mathbb{R}$). Note that the other boxes \eqref{eq:box2} also admit such inputs on one of their side.


\begin{thebibliography}{99}
\bibitem{BellB}
J. S. Bell, \emph{Speakable and Unspeakable in Quantum Mechanics} (Cambridge University Press, 1987).
\bibitem{Aspect99}
A. Aspect, Nature {\bf 398}, 189 (1999).
\bibitem{Toner}
B. F. Toner and D. Bacon, Phys. Rev. Lett. {\bf 91}, 187904 (2003).
\bibitem{Barrett04}
J. Barrett, N. Linden, S. Massar, S. Pironio, S. Popescu, D. Roberts, Phys. Rev. A {\bf 71}, 022101 (2005).
\bibitem{CGMP}
N.J. Cerf, N. Gisin, S. Massar and S. Popescu, Phys. Rev. Lett. {\bf 94}, 220403 (2005).
\bibitem{Brunner08}
N. Brunner, N. Gisin, S. Popescu and V. Scarani, Phys. Rev. A {\bf 78}, 052111 (2008).
\bibitem{Tessier05}
T. E. Tessier, C. M. Caves, I. H. Deutsch and B. Eastin, Phys. Rev. A {\bf 72}, 032305 (2005).
\bibitem{Barrett07}
J. Barrett, C. M. Caves, B. Eastin, M. B. Elliott and S. Pironio, Phys. Rev. A {\bf 75}, 012103 (2007).
\bibitem{BroadBentTapp}
A. Broadbent, P. R. Chouha and A. Tapp, arXiv:0810.0259.
\bibitem{Popescu94}
S. Popescu and D. Rohrlich, Found. Phys. {\bf 24}, 379 (1994).
\bibitem{Yao82}
A. C. C. Yao, in \emph{23rd Annual Symposium on Foundations of
Computer Science, Chicago, 1982}, (IEEE, New York, 1982), p.160.
\bibitem{Collins02}
D. Collins, N. Gisin, S. Popescu, D. Roberts and V. Scarani, Phys. Rev. Lett. {\bf 88}, 170405 (2002).
\bibitem{Seevinck02}
M. Seevinck and G. Svetlichny, Phys. Rev. Lett. {\bf 89}, 060401 (2002)
\bibitem{Bancal09}
J.-D. Bancal, C. Branciard, N. Gisin, S. Pironio, Phys. Rev. Lett. {\bf 103}, 090503 (2009).
\bibitem{svetlichny}
G. Svetlichny, Phys. Rev. D {\bf 35}, 3066 (1987).
\bibitem{Barrett05}
J. Barrett and S. Pironio, Phys. Rev. Lett. {\bf 95}, 140401 (2005).
\bibitem{Bancal09b}
J.-D. Bancal, N. Gisin ans S. Pironio, J. Phys. A: Math. Theor. {\bf 43}, 385303 (2010).
\bibitem{kushilevitz_nisan}
E. Kushilevitz, N. Nisan, \emph{Communication complexity} (Cambridge University Press, 1997).
\end{thebibliography}
\end{document}